\documentclass[journal]{IEEEtran}
\usepackage{color}
\usepackage{graphicx}
\usepackage{subfigure}
\usepackage{cite}
\usepackage{amsmath} 
\usepackage{amssymb}  

\ifCLASSINFOpdf
\else
\fi
\begin{document}
%
\title{Mission Oriented Miniature Fixed-wing UAV Swarms: A Multi-layered and Distributed Architecture}
%
%
%

\author{Zhihong Liu,
       Xiangke Wang,
       Lincheng Shen,
       Shulong Zhao,
       Yirui Cong,
       Jie Li,
       Dong Yin,
       Shengde Jia,
       Xiaojia Xiang
       
\thanks{Z. Liu, X. Wang, L. Shen, S. Zhao, Y. Cong, J. Li, D. Yin, S. Jia and X. Xiang are with the College of Mechatronics and Automation, National University of Defense Technology, Changsha,
HN 410073, China (e-mail: \{zhliu, xkwang, lcshen, jaymaths, congyirui11, leonlee2009, yindong, jia.shde, xjxiang\}@nudt.edu.cn).}
}

\maketitle

\begin{abstract}

UAV swarms have triggered wide concern due to their potential application values in recent years. 
While there are studies proposed in terms of the architecture design for UAV swarms, two main challenges still exist: (1) Scalability, supporting a large scale of vehicles; (2) Versatility, integrating diversified missions. To this end, a multi-layered and distributed architecture for mission oriented miniature fixed-wing UAV swarms is presented in this paper. The proposed architecture is built on the concept of modularity. It divides the overall system to five layers: low-level control, high-level control, coordination, communication and human interaction layers, and many modules that can be viewed as black boxes with interfaces of inputs and outputs. In this way, not only the complexity of developing a large system can be reduced, but also the versatility of supporting diversified missions can be ensured. Furthermore, the proposed architecture is fully distributed that each UAV performs the decision-making procedure autonomously so as to achieve better scalability. Moreover, different kinds of aerial platforms can be feasibly extended by using the control allocation matrices and the integrated hardware box.
A prototype swarm system based on the proposed architecture is built and the proposed architecture is evaluated through field experiments with a scale of 21 fixed-wing UAVs. 
Particularly, to the best of our knowledge, this paper is the first work which successfully demonstrates formation flight, target recognition and tracking missions within an integrated architecture for fixed-wing UAV swarms through field experiments. 


\end{abstract}


%
\IEEEpeerreviewmaketitle

\section{Introduction}

Due to the advantages in flexibility, cost and 
environmental adaptability, unmanned aerial vehicles (UAVs) have created tremendous application potential and have been increasingly investigated in recent years. 
In particular, UAVs are widely used in the areas such as reconnaissance, surveillance, plant protection and disaster rescue. However, with the advance in coordination technology, the limitations of using single UAV to operate missions become more and more apparent. 
UAV swarms, consequently, have attracted much attention. Through coordination between members, UAV swarms can share the resources of the whole system and can work as a team cooperatively.
In this way, UAV swarms can be more competent for large complex missions.

In order to increase the level of autonomy for UAV swarms, a large amount of studies have been proposed in the area of UAV swarming over the past few years. Some proposals focus on the flocking control
\cite{Preiss2017Crazyswarm,de2017circular}.  Some proposals concentrate on 
the mission planing and decision making \cite{bai2018integrated,jin2018distributed}. Some proposals study the target recognition and tracking \cite{8331947,7330001}.  However, few research is revealed in the perspective of the architecture which plays an importance role at the system design and implementation.


In particular, Sanchez-Lopez et al. \cite{Sanchez2015A,sanchez2016aerostack} propose an open-source architecture named by AeroStack for multi-UAV systems. This architecture follows a hybird reactive/deliberative paradigm and includes five layers, i.e., reactive, executive, deliberative, reflective and social layers. Whereas AeroStack deploys the time-critical control (e.g. attitude control, actuator control) on a non-real-time system, which may fail to satisfy with the real-time requirements for high speed UAVs. Grabe et al. \cite{Grabe2013The} propose Telekyb, an end-to-end control framework for controlling heterogeneous UAVs. Although it allows coordination control of multiple UAVs, its scalability is limited. This is because in Telekyb, 
the high-level control (e.g. mission planning) operates on the ground rather than on-board. Boskovic et al. \cite{boskovic2009collaborative} propose CoMPACT, a six-layered hierarchical architecture for controlling swarms of UAVs. The main advantage of CoMPACT is that it effectively combines top-level mission planning and decision making with dynamic re-assignment, reactive motion planning and emergent biologically-inspired swarm behaviors. 
Nevertheless, CoMPACT splits the mission execution to levels of mission, function, team, platoon, UAV, and each level requires a manager that cooperates with other UAVs in corresponding level. This may increase the burden of the task management. Note that these aforementioned works are evaluated by simulations or experiments for quadrotors, and no field experiments for fixed-wing UAVs are demonstrated. Comparatively, Chung et al. \cite{chung2016live} propose a swarm system and demonstrate live-fly field experiments with up to 50 fixed-wing UAVs. However, this work mainly focuses on the system design for UAV flocking including the autonomous launch, flight, and landing. The collective behaviors and mission coordination are not included in this swarm system. 





Although this field of research has brought important contributions, there are mainly two remaining challenges: 1) \textbf{Scalability}. Most of the related work are evaluated by experiments of small scales (i.e. two to five). It is known that with the scale increases, system designs are more challenging both theoretically and practically. A scalable architecture that can support a large scale of UAVs is needed. 2) \textbf{Versatility}. Existing solutions mainly focus on specified problems or applications. Few to achieve an integrated framework for multiple uses. However, high degrees of autonomy for UAV swarms requires the ability of supporting multiple and heterogeneous applications (e.g., flocking, target recognition and tracking). 
Therefore, an architecture that integrates diversified functions and missions is desired. 




According to these needs and issues, we present a multi-layered and distributed architecture for mission oriented fixed-wing UAV swarms. 
Compared to other architectures and frameworks, there are three main contributions.

\begin{itemize}

\item Firstly, the proposed architecture is built on the concept of modularity and divides the overall swarm system to multiple layers and many modules. 
It allows each module focus on its own design and abstracts away the details of other modules, which facilitates the implementation and the extension for developers.
As a result, not only the difficulty of developing a large system can be reduced, but also the versatility of supporting diversified missions can be ensured.

\item Secondly, the proposed architecture is fully distributed and each UAV performs the decision-making procedure (abides by Observe-Orient-Decide-Act, OODA) autonomously. By this means, it removes the dependence of central controller for mission coordination and brings better scalability to UAV swarms. 

\item Thirdly, the proposed architecture is not restricted to specified kinds of aerial platforms. Through introducing control allocation matrices and the platform-independent integrated hardware box, different kinds of aerial platforms can be feasibly extended to the swarm system.  We have accomplished flight experimentations of a swarm with heterogeneous aerial platforms including fixed-wing and tilt-rotor aircrafts.



\end{itemize}

Through field experiments with a scale of 21 fixed-wing UAVs, we evaluate the scalability and versatility of the proposed architecture. Several coordinative missions such as formation flights, target recognition and tracking are demonstrated. Particularly, to the best of our knowledge, this paper is the first work to successfully demonstrate formation flight, target recognition and tracking missions within an integrated architecture for fixed-wing UAV swarms through field experiments. Besides, the experimental results also show that the launch rate of the prototype system based on the proposed architecture outperforms the state-of-the-art work.

The rest of the paper is structured as follows. Section~\ref{sec:arch}
presents the overview of the proposed architecture. The design of the low-level control layer is presented in Section \ref{sec:lowlevel}. Section~\ref{sec:highlevel} describes the high-level control layer. The communication layer and the human interaction layer are elaborated in Section~\ref{sec:comm} and Section~\ref{sec:gcs}, respectively. Section~\ref{sec:results} reports the results of the field experiments. And Section \ref{sec:conclusion} concludes the paper and indicate future research directions.

\begin{figure*}[t]
	\centering
	\includegraphics[width=0.9\textwidth]{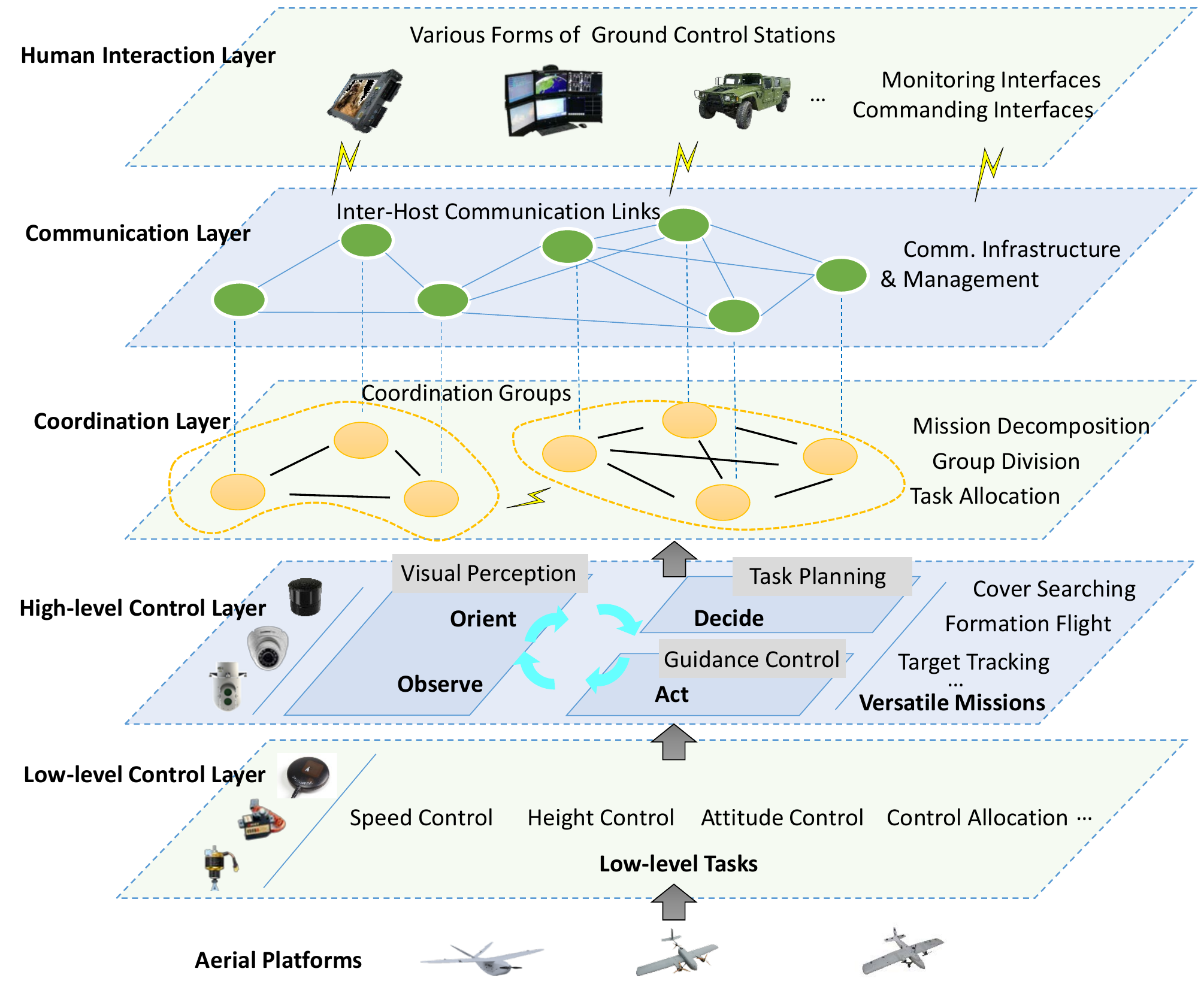}
	\caption{System architecture.}
	\label{fig:systemArch}       
\end{figure*}

\section{System Architecture}\label{sec:arch}

In order to maximize the superiority of UAV swarms, there are four key capabilities that an UAV swarm system needs to obtain. First, to support a large scale of UAVs. Second, to handle diversified missions. Third, to coordinate with other UAVs among the swarm efficiently. Fourth, to support heterogeneous aerial platforms. For the purpose of acquiring these capabilities, we have designed a multi-layered and distributed architecture for organizing the the fixed-wing UAV swarm system's functional modules and subsystems.
The full system architecture is outlined in Fig.~\ref{fig:systemArch}. It mainly consists of five layers: low-level control, high-level control, coordination, communication and human interaction layers. 

The low-level control layer is deployed on an embedded real-time operating system which guarantees minimal system interrupt latency and thread switching latency. Hence, it is qualified for the work of flight control (e.g., attitude control and actuator control). 
The high-level  control layer 
is deployed on a high-performance processing board which makes it possible to run computation intensive tasks, such as visual perception, task planning and guidance control, on-board. And the tasks performed in this layer abide by Observe-Orient-Decide-Act (OODA) procedure.  
The coordination layer encapsulates the functions in terms of the negotiation (e.g., task allocation) among UAVs for cooperative missions. Through the coordination layer, each UAV can negotiate with other UAVs to obtain free-conflict solutions. Like the high-level control layer, this module is also deployed on a high-performance processing board. 
The communication layer manages the message transmission among all the UAVs and the ground control systems. It includes the design of the  communication infrastructure (from the perspective of hardware) and the communication management (in terms of software).   
The human interaction layer is deployed on the ground and provides interfaces for visualizing the situation including the UAV status, the sensed data and the geographical environment. And it also offers interfaces for operators to command the UAV swarm system.

Through slicing the swarm system to five layers with specified functionalities, this architecture reduces
the complexity of developing a large system. Moreover, the proposed architecture divides the overall system to many modules that can be viewed as black boxes with interfaces of inputs and outputs. 
In this way, each module focuses on its own design and abstracts away the details of other modules, which facilitates the implementation and the extension for developers. 

The proposed architecture is fully distributed and brings better scalability. Each UAV performs the decision-making procedure autonomously. In this way, it removes the dependence of central controller for mission coordination. Moreover, this architecture dynamically divides the UAV swarm to individual coordination groups according to  the mission requirements and the communication availability, which can hold the scale of the states maintained on each UAV for making decision as the number of UAVs increases. Therefore, the scalability of the swarm system can be significantly improved.

In order to satisfy the timing and computing requirements for the controlling of the swarm system in different levels, the proposed architecture leverages two kinds of processing boards.  One uses low-power micro controller unit and is installed with embedded real-time operating system. In this processing boards, the tasks with strong real-time requirement (e.g. attitude control and actuator control ) can be deployed, which is respect to the low-level control layer in the proposed architecture. The other uses the high-performance micro processing unit and is installed with time-sharing operating system. In this processing boards, the computation intensive tasks (e.g. target recognition and mission planing) can be deployed, which is corresponding to the high-level control and coordination layers. Therefore, this design of the two level processing not only compensates insufficient computing capacities for the real-time platform, but also brings more flexibility for implementing the high-level algorithms.

Note that the proposed system architecture is not restricted to specified kinds of aerial platforms. It is true that different platforms may have different configurations such as payloads, propulsion mechanisms, shapes and weights. It is also known that the same flight control signals~(e.g. speed, attitude, altitude, etc.) produce different actuator control outputs for aerial platforms with different configurations. By introducing the control allocation matrices to differentiate the aerial platforms, the swarm system can convert the low level control signals to compatible actuator control outputs according to the configurations of platforms dynamically.  Moreover, for the purpose of designing a lightweight and miniaturized system, the proposed architecture integrates the on-board hardwares (e.g. the processing boards, perceptional devices, communication payloads, circuitry and cooling devices) into a compact box. This box is loosely-coupled with the aerial platform. As a result, different kinds of aerial platforms can be feasibly extended to our swarm systems by installing the integrated hardware box. 
Based on this system architecture, we have accomplished flight experimentations of a swarm with hybrid aerial platforms including fixed-wing and tilt-rotor aircrafts. In the following subsections, we will provide the details of each component of the proposed system architecture.


\section{The Low-level Control Layer}\label{sec:lowlevel}

The low-level control layer is in charge of the flight control for the UAVs in swarms, which provides each UAV with the ability of accurate flight and adaptation to the complex environment. In this layer, on-board sensors, such as accelerometers, magnetic compasses, and gyroscope, are attached and able to provide the current position and attitude information of the UAV in a timely manner. And this layer accepts the command references of the upper layer and converts to the desired attitude. After obtaining the appropriate pulse-width modulation (PWM) output according to the attitude instruction and the current state of the UAV, the control signal is transmitted to actuators (aileron, elevator, rudder and throttle).

The fixed-wing UAV is commonly regarded as a six-degree-of-freedom (DOF) rigid body,   and it is well known that the dynamic characteristics and control principle of fixed-wing UAVs are quite different from those of quadrotors and helicopters \cite{castillo2006modelling}. For example, the fixed-wing UAVs are required to maintain a minimum airspeed  to produce enough lift force,  resulting in lacking hovering capability. 
Further, the dynamical model of a fixed-wing UAV is characterized by air-operated complexity, manipulative coupling and controllable underactuation. It is hard to establish the accurate dynamical of miniatured model for cost reasons.  
In addition, cross-coupling dynamic characteristics are generally demonstrated for fixed-wing UAVs, which makes their flight performance vulnerable to both external disturbances and inner effects \cite{liu2016disturbance}.  
Overall, the accurate flight control scheme of  the low-cost miniature fixed-wing UAV is of importance and very challenging.  Hence, the low-level control layer plays a pivotal role in supporting the whole UAV swarm system.

In our work, there are mainly three aspects in the low-level control layer: speed and height control, attitude control and control allocation. The accuracy of the heading control ensures that the nose of the vehicle can follow the desired heading angle within an acceptable range. The control accuracy of the speed control ensures the coordination of the UAV in space. Attitude (pitch and roll) control is the most critical part of the flight controller. Its control frequency is usually several times that of the upper layer, and its performance directly affects the safety and stability of the vehicle. 

\subsubsection{Speed and height control}
Fixed-wing aircraft rely on wings to generate lift, so the forward flight speed of an aircraft  is primarily related to its ability to drive: 
\begin{eqnarray}\label{1}
\frac{dV}{dT}=\frac{T-D}{m}-g\sin\gamma,
\end{eqnarray}
where, $T$ indicates the thrust of the engine, and $D$ expresses the resistance.
From the perspective of capacity conservation, there is a coupling relationship between aircraft speed and altitude:
\begin{eqnarray}\label{1}
E_T=E_D+E_S=\frac{1}{2}mV^2+mgh,
\end{eqnarray}
where, $V$ and $h$ represent the speed and altitude of the vehicle, respectively. When controlling the speed of the aircraft, it is necessary to consider both the thrust of the engine and the pitch angle of the aircraft. Due to the coupling relationship between the speed and height of the fixed-wing UAV, simply adjusting the drive of the vehicle cannot fully control the speed. Here we use a fuzzy controller that takes the altitude and speed of the UAV as inputs and takes the pitch angle and throttle as control outputs. The built-in expert logic relationship is used to improve the corresponding characteristics of the speed and height control.

\subsubsection{Attitude control}
The most important thing to consider when the aircraft can stably fly is the balance between the lateral stability surface and the vertical stability surface. The requirements of roll angle and pitch angle control are fast, stable, and easy to implement. Many methods can effectively achieve attitude stability such as \cite{Zhao2017Curved}. In general, the adjustment of the course angle of a fixed-wing aircraft can be seen as a circular turn: 
\begin{eqnarray}\label{1}
\dot\chi & = & \frac{g}{V_g}\tan\phi_c,
\end{eqnarray}
where, $\dot\chi$ represents the heading angle and $\phi_c$ is the desired control input of roll angle. Adjusting the nose of the body mainly rely on the roll angle, that is, the change of the roll angle brings changes in the course angle. There are many methods used for heading control \cite{Zhang2017Autonomous,Zhao2017Model}, and what they have in common is easy to implement and robust to disturbance.

\subsubsection{Control Allocation}

The same flight control signal (e.g. attitude control) produces different actuator control outputs for vehicles of different configurations. Due to the different aerodynamic layout and configuration, such as conventional configurations, delta wing, flying wing, double tail, v-tail, etc., the vehicle's actuator outputs are completely inconsistent. In order to be able to adapt to more vehicle platforms, we introduce a concept of the control allocation matrix and ensure that different configurations correspond to different distribution matrices \cite{Pedro2017PI}.

The design of the distribution matrix is based on parameters such as the size, weight and actuator performance of the platforms. With the distribution matrix, we can convert the control signals of the low-level control (speed, altitude, attitude, etc.)  into compatible control outputs of the actuators for different platforms. 



\section{The High-level Control Layer}\label{sec:highlevel}
The high-level control layer concentrates on the tasks such as visual perception, mission planning, guidance control, etc. It 
follows the Observe-Orient-Decide-Act loop for realizing swarm autonomously. More specifically, an electro-optical device is attached to this layer and the visual perceptional processing module, which provides the information related to the targets and obstacles, is included in the layer. This represents the observe and orient procedures. In addition, the mission planning module is deployed for producing task plans that can accomplish the user demanding missions. This stands for the decide procedure. Besides, the guidance control that intends to guide the UAVs to reach the desired point in coordination with other UAVs is included, which is implied for the act procedure. The high-level control layer is on top of the low-level control layer but under the coordination layer. It leverages the upper layer to negotiate with other UAVs and produces the guidance control commanding references to the lower layer. 

\subsection{Visual Perception}

Perception for UAVs intends to become aware of the current state of itself and the environment through on-board sensors . Due to higher requirements of the update rate, the awareness of its current state (e.g. attitude, velocity and airspeed) is deployed on the low-level controller, where the sensors such as IMU, gyroscope and compass are attached. Here, we consider to use the vision and range devices. And the visual perception module is responsible of high-level perceptional processing including target recognition, target localization, obstacle detection and situation awareness. 

\textbf{Target recognition} has been studied for years and plenties of proven solutions have been proposed. Recently, this technique has been widely used in unmanned systems \cite{kechagias2018new}, e.g., life search in disaster rescue, criminal chase in urban area, etc. By attaching cameras or other imaging devices (e.g. infrared and hyper-spectrum), image or video stream can be obtained continuously. Identifying the interested objects from the incoming image or video data in real-time, and thereby providing detection results to other decision module timely becomes an essential procedure of accomplishing the missions.  

\textbf{Target localization} is trying to localize ground-based objects based on image data from UAVs. By using the target recognition results, the pixel location of the target in an image can be obtained. Hence, according to the pixel location, the UAV's attitude and position, and the camera angles, the target localization in world coordinates can be estimated. Besides, in order to improve the accuracy of estimating the status of targets (e.g. positions, velocities), cooperative target status estimation algorithm using multiple UAVs can be included~\cite{minaeian2016vision}. 

\textbf{Obstacle detection} and avoidance for UAVs is an anticipated requirement for autonomous flights, especially in low-altitude maneuvers surrounding with trees, buildings and other structures. In general, range sensors (e.g. LiDAR, infrared and ultrasonic) are exploited to measure the distance between the UAV and obstacles \cite{gageik2015obstacle}. Based on this, the safety area around the UAV can be computed. Other than this category of approaches using active sensors, the vision-based obstacle detection approaches perceive obstacles through passive sensors such as cameras. However, intensive-computation operations such as image feature tracking and three dimensional world information constructing are needed in these approaches. 


Note that the perception module is not restricted to the tasks introduced above. Thanks to the modularity inhered from ROS, other perception tasks such as 
situation understanding and simultaneous localization and mapping (SLAM) algorithms can also be easily included.

\subsection{Task Planning}

\begin{figure}
	\centering
	\includegraphics[width=0.5\textwidth]{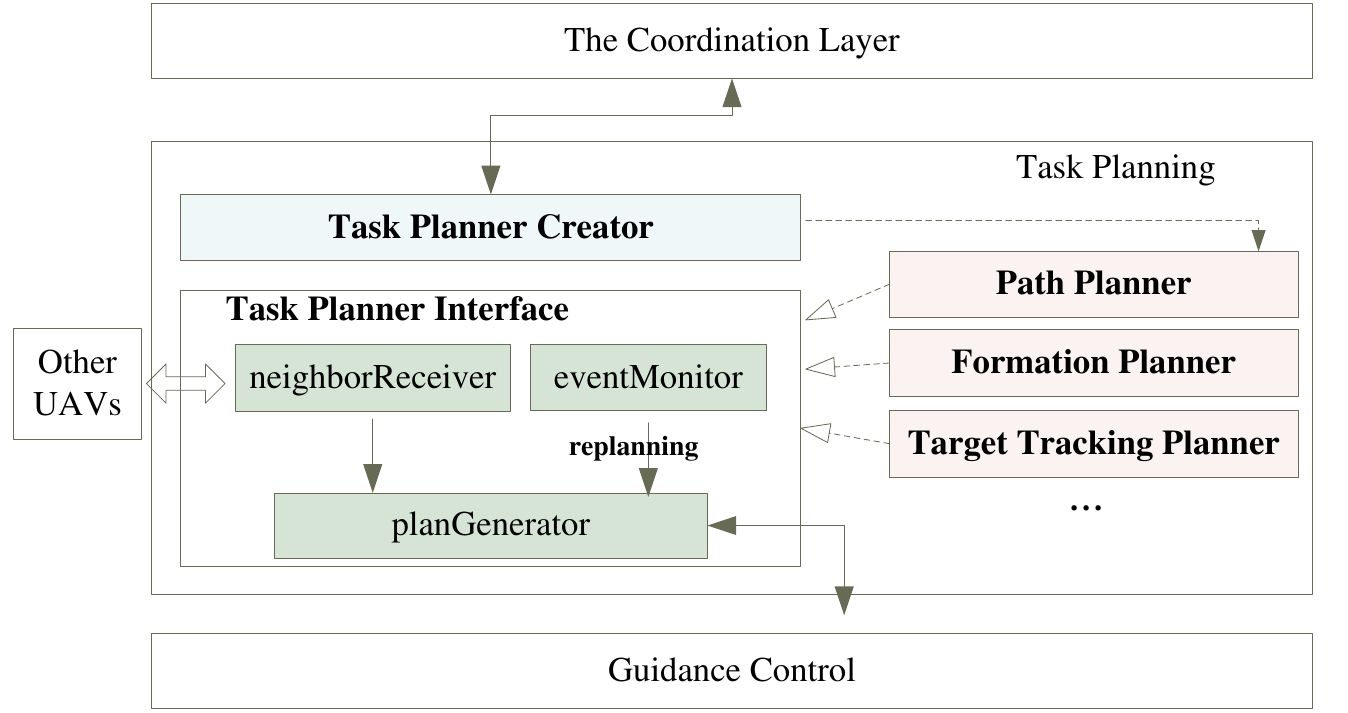}
	\caption{Task Planning Architecture}
	\label{fig:missioinPlan}       
\end{figure}

The Task planning module produces task plans that satisfies with the requirements for the commanding missions while also abiding by constrains such as UAV's payload, endurance and airspace regulation. This is critical for maximizing the capabilities of UAV swarms as well as the quality of the mission completion. 
Generally speaking, there are two perspectives of mission planning for UAV swarms. The first is to allocate tasks\footnote{A mission can be decomposed into multiple tasks} among multiple UAVs and schedules the tasks for each UAV in an proper sequence. The second is to generate a detailed plan for each task in the sequence. In the proposed architecture, the first perspective is deployed in the coordination layer (see Section \ref{sec:coor}), and the task planning module is in charge of later perspective.  

In order to improve the scalability of the task planning module, we apply the Factory design pattern\cite{morton1992object}. In this way, planners that have different purposes but abide by the unified interface can be extended easily. The architecture of the task planning scheme is shown in Fig. \ref{fig:missioinPlan}. More specifically, it consists of three types of components: the Task Planner Interface, the Task Planner Factory and different kinds of planners. The Task Planner Interface declares the methods that each instance of the planners need to implement. More specifically, the neighborReceiver receives the state of the neighbor UAVs and the planGenerator develops a coordination plan based on the state of neighbor UAVs. Besides, the eventMonitor monitors the event and status at runtime and re-plan dynamically once some predefined conditions are triggered. The instances of the planners are the actual planners which provide the implementation to the methods declared in the Task Planner Interface. The Task Planner Factory constructs the actual planners according to the requests from the coordination layer. In our proposed architecture, we includes path planner, formation planner and target tracking planner described below.

\begin{itemize}
	\item Path Planner generates collision-free routes that can be satisfied with the task requirements (e.g. destination, threat avoidance, path length) while taking into account the geometric and physical constraints. In order to generate routes for each cooperative UAV in the swarm, this planner gathers the status of neighbor UAVs and calculates the optimized routes in a distributed manner. Besides,  re-planning is triggered automatically while emergencies is detected. Note that path planning problems have been actively studied and many solutions can be included here.
	
	\item Formation Planner configures the formation patterns of a UAV swarm according to mission requirement and geographical conditions, e.g., straight line, triangle pattern, victory pattern (as shown in Fig. \ref{fig:reconfig}). After that, it is also responsible for generating the trajectory for each leader UAV and the relative location to its leader UAV for each follower UAV. Besides, the formations can also be dynamically reconfigured as required, e.g., while crossing a narrow valley, the formation pattern of the UAV swarm can be changed to straight line. 
	
	\item Target tracking planner is in charge of generating routes for arriving at and covering the target area. That is because the area of the moving target is provided with the mission, rather than a precise location. UAV swarms need to search the target in the corresponding area. It is noted that tracking targets in the mission planning level cannot handle the high uncertainty of the target's moving state. In our design, this task is implemented in the guidance control level (see Section \ref{sec:guidance}), which is also consistence with the state-of-the-art solution\cite{oh2015coordinated,xiong2017guidance}.  
	
\end{itemize}

\subsection{Guidance Control}\label{sec:guidance}
The guidance control module intends to guide the hosted UAV to reach the desired points or follow the command references produced by the task planning module. More specifically, it uses the UAV status from the low-level control layer and other UAVs as feedback, and produces the  control command references such as desired yaw, speed and height for the hosted UAV.

Similar to the task planning module, we apply the Factory pattern for extending different kinds of guidance control algorithms, e.g., path following, formation control, target tracking. We can select corresponding kinds of guidance control algorithms according to the task plan produced by the task planning module. Besides, with respect to the same kinds of guidance control, our proposed architecture can extend different control laws by implementing different algorithms and adjust them dynamically according to the mission requirements. Also, even for the same guidance control algorithms, control parameters can be adjusted by configuration or user commands. The guidance control algorithms included in our architecture are as follow.

\textbf{Path following:}
After completing the path planning, each UAV will need to follow the path accordingly. No matter it is coordinated or singular  path following, the core goal of each vehicle is to follow a desired path 
so that the tracking error remains within an acceptable range. 
One of the most basic capabilities of path following is to effectively resist wind disturbances and keep the aircraft on the desired path.

Due to the coupling of the height and speed of the fixed-wing aircraft, the adjustment of the position is mainly achieved by its speed and heading. For ease of analysis, the problem of path following is usually decoupled into two-dimensional path following and height control.  By designing a vector field around the desired path, 
the orientation of the aircraft head is determined in the vector field according to the distance from the current position of the aircraft to the desired path, and the tracking error is ensured asymptotically converging to zero~\cite{zhao2018integrating}. 
Admittedly, other common path following control methods, such as PLOS\cite{kothari2010suboptimal}, NLGL\cite{park2007performance}, and  LQR-based path following\cite{ratnoo2011adaptive} , can also be adopted here.

\textbf{Formation control:}
The formation control intends to control a group of UAVs flying in formation cooperatively. And the formation pattern should be preserved during maneuvers such as heading change and speed change. Over the past decades, many formation control approaches have been proposed such as consensus-based approach\cite{wang2016multi}, leader-follower approach \cite{dehghani2016communication}, behavior based approach \cite{qiu2015multiple} and virtual structure approach \cite{askari2013uav}. In our work, we  adopt a hybrid formation control approaches. For the leader UAVs,  we use coordinated path following control; with respect to the follower UAVs, we use leader-follower coordinated control.  

By using the coordinated path following techniques, the guidance control module receives paths that  are parameterized according to the desired formation for the hosted UAV. More specifically, there are some special waypoints in which the leader UAVs are required to arrive at the same time. Thus, the control law should not only ensure that each path following error converges to zero, but also achieve finite-time stability and consensus for desired speed. As a result, a desired inter-vehicle formation for leader UAVs can be achieved.

While the leader UAVs fly as the mission required, the follower UAVs follow the corresponding leader UAV and try to form the formation. More specifically, each follower UAV firstly generates an induce route according to the desired distance between its current position and its desired position. After that, each follower UAV will be guided by the path following guidance law with velocity adaption.

\textbf{Target tracking:}
The guidance control for target tracking is responsible for guiding the UAV to fly around the targets 
so that the targets remain in the UAV's detection range. More specifically, the UAV first follows a path that can arrive at and cover the target area. Once the target is detected by the visual-perceptional module, the UAV fly a circular orbit around the targets and keep a constance distance with the targets. The first aspect of the work can be done by the path following guidance control. And the guidance control for target tracking is mainly implemented by the vector field methods \cite{lim2013standoff,chen2013uav}. With respect to single-vehicle tracking, this kind of methods builds a Lyapunov vector field for controlling the heading of the UAV in order to guide the UAV to fly a circular orbit around the target with a specified radius. When it comes to multiple-vehicle tracking, this kind of methods builds an additional Lyapunov vector field for controlling both the desired speed and heading of the UAV in order to ensure the inter-vehicle angular spacing, thereby preventing these UAVs from collision and achieving multiple viewing angles for surveillance.


Note that  the obstacle and collision avoidance functionality is also included in this module. More specifically, the condition of performing obstacle and collision avoidance actions can be triggered any time during the mission execution. Once the condition is triggered, the obstacle and collision avoidance algorithm produces guidance command references for the host UAV in order to avoid the detected obstacle and collision with other UAVs. This has higher priority than the mission executions.
While safety condition is satisfied, UAVs continue the previous mission.

\subsection{Supervision}
Supervision module is in charge of maintaining the system healthy inspection, allowing or disallowing operations based on mission requirements, 
recording the flight logs, etc. This is an essential module for ensuring the stability and proper functioning of the system. In particular, it consists of Commander and Logger sub-modules that are explained below. 

\textbf{Commander} inspects the healthy status of all modules, maintains a state machine in terms of the system level (e.g. current mode), and allows or prohibits actions according to its current state or mission requirements. In addition, commander is in charge of dispatching all the commands given by the ground station. Different commands may need different operator. For example, commands of mission execution class such as target tracking and formation flight need to be dispatched to the mission planner; and commands of system management class such as the predefined trajectories loading and configuration changing need to be dispatched to the logging and storage.

\textbf{Logger} logs the state of the system including the state transaction, trigged events, flight data and custom logs. Other than this, the perceived images or videos can be stored on the platform selectively. This is very helpful for developers to analyze the UAV's behaviors and performance over the entire process. Besides, configuration files, parameter lists, predefined waypoint lists are also stored on-board in order to fast fetch for on-board modules.  

\begin{figure}
	\centering
	\includegraphics[width=0.5\textwidth]{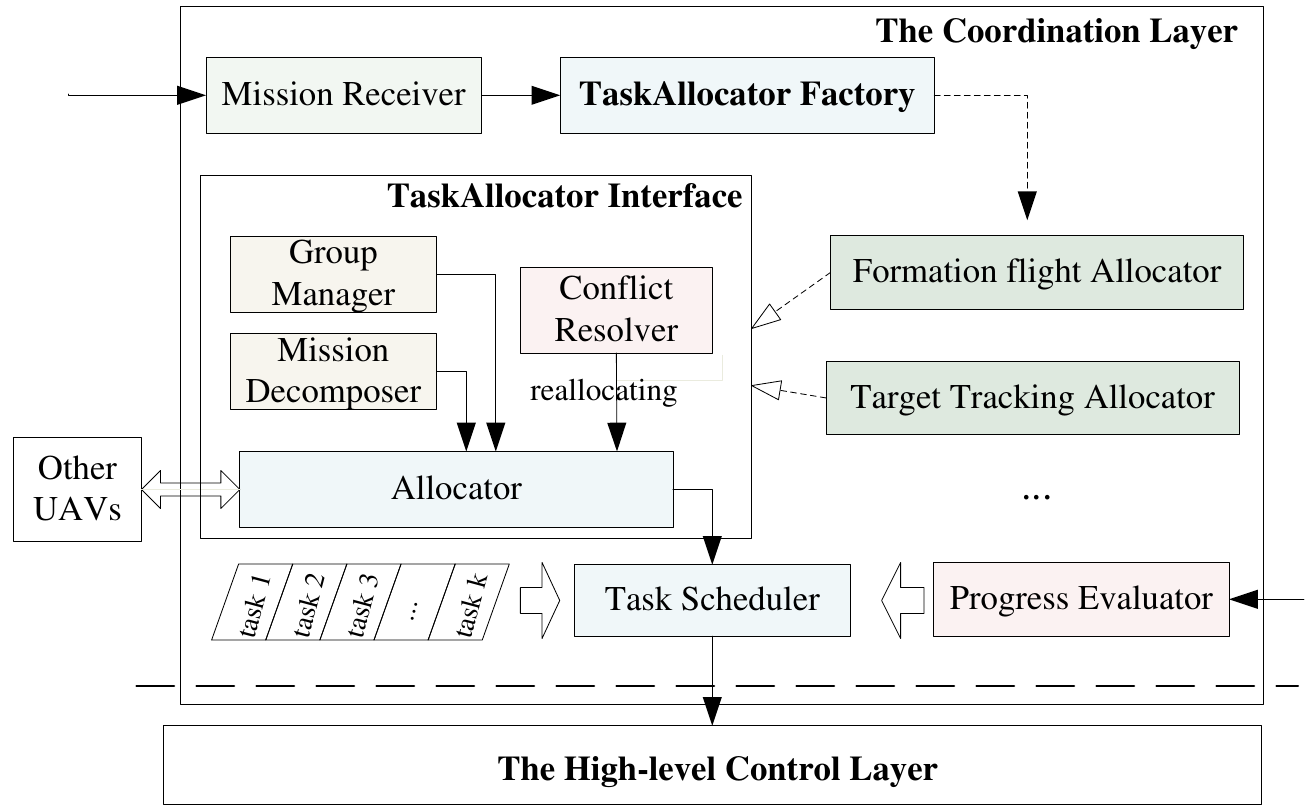}
	\caption{The overview of the coordination layer.}
	\label{fig:coordination}       
\end{figure}

\section{The coordination layer}\label{sec:coor}

The coordination layer is in charge of the tasks related to negotiation (e.g., task allocation) among UAVs for mission coordination. More specifically, it is responsible of  dynamically dividing the UAV swarm to individual coordination groups according to the mission requirements and the communication availability, which can restrict the scale of the states maintained on each UAV for making decision as the number of UAVs increases. Besides, it is also in charge of decomposing the mission to tasks and the distributing the tasks among the UAVs in the coordination group. Note that there is a task planner in the high-level control layer being in charge of generating the detailed plan for the obtained tasks correspondingly. Moreover, it is responsible of reallocating the tasks as well as the task sequences while conflict is detected. 

It is noted that for different missions, the negotiation work among the UAV swarms should be different, for example, the objective of task allocation for formation flight and target tracking is to decide which UAVs are the leaders or the follower and which targets should be tracked by which UAVs, respectively. Therefore, similar to the task planning module, we apply the Factory pattern to include different task allocators. The design of the coordination layer are shown in Fig. \ref{fig:coordination}.

Once a mission is received, the TaskAllocator Factory launches an associated task allocator according to the mission type. In each task allocator, it consists of the Group Manager, the Mission Decomposer, the Allocator and the Conflict Resolver. More specifically, the Group Manager decides the coordination group based on the mission requirements and communication availability. Here, clustering algorithms can be used to divide the groups. The Mission Decomposer decomposes the mission to tasks according to the preset configuration. The allocator allocates the tasks as well as the task sequence among UAVs in the coordination group. Many task allocation approaches such as market-based \cite{choi2009consensus} and optimization-based \cite{wu2018distributed} mechanisms can be used here. The conflict resolver detects the potential conflict and triggers reallocation while needed. After the task sequence is determined, the Task Scheduler schedules the tasks one by one in the task sequence queue. Note that there are two kinds of tasks in the perspective of scheduling: blocking and non-blocking tasks. When the blocking task is completed, the sequential task can be launched; whereas the non-blocking tasks is launched, the sequential task can be launched.

\begin{figure}
	\includegraphics[width=0.5\textwidth]{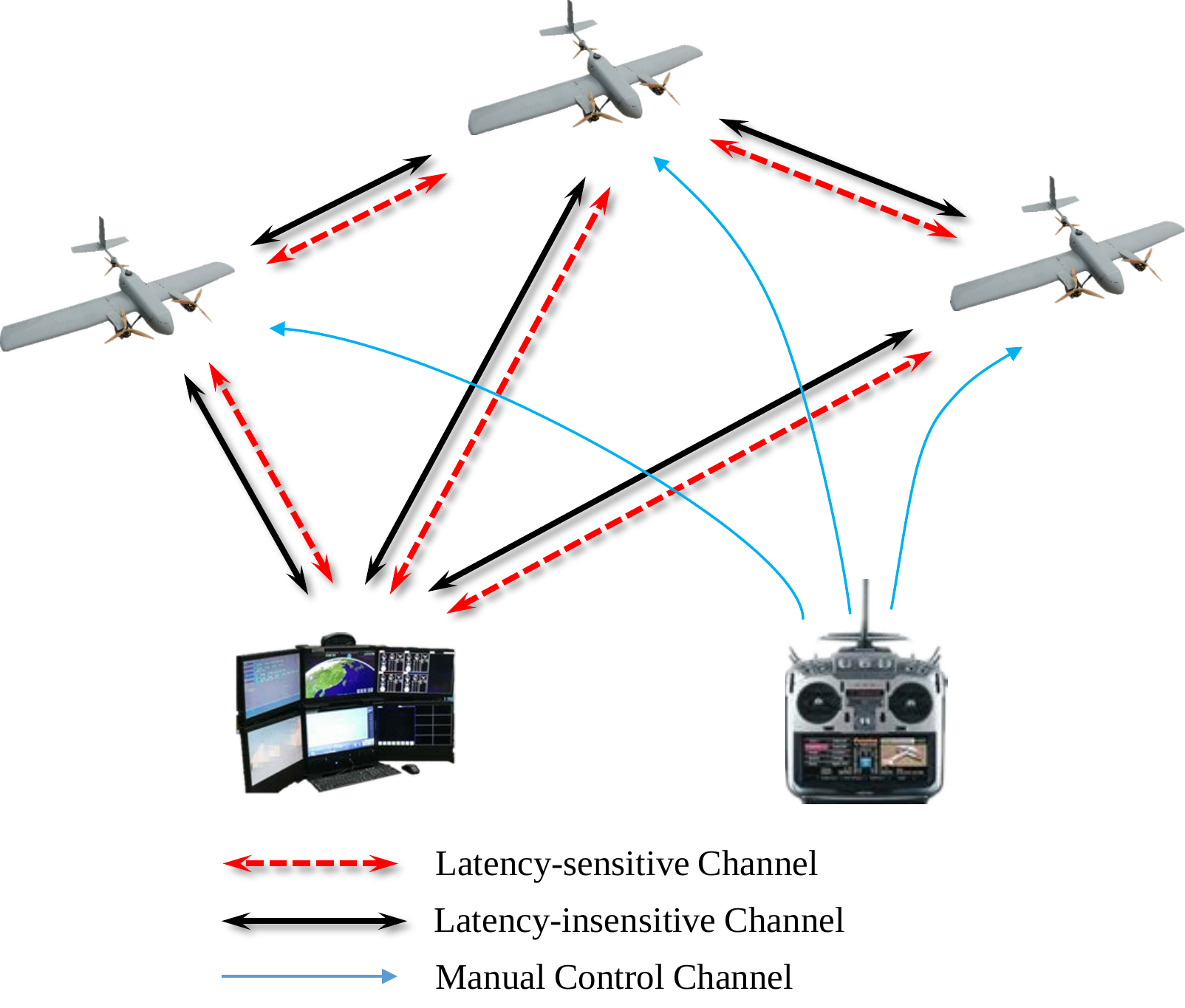}
	\caption{Different channels of the communication infrastructure.}
	\label{fig:comm}       
\end{figure}

\section{The communication layer}\label{sec:comm}

The communication layer is in charge of the messages transmission among all the UAVs and the ground control station, which is a critical desire for a highly autonomous and cooperative swarm system.
This is because the essential tasks including command execution, coordination messages and sensed data feedback (e.g. position, image, video) largely rely on the quality of communications.
Furthermore, the communication system should include the following features:
\begin{itemize}
	\item   The communication system not only needs the air-to-ground and ground-to-air links to support the connectivity between the ground station and UAVs, but also requires the air-to-air links to help the information exchanges between different UAVs.
	\item   It is necessary to support different types of transmissions in the communication system which correspond to different quality of services.
	For example, the video (presenting in the ground station) and position information (supporting formation flight) have different data rate and latency requirements.
	\item It is scalable to the number of UAVs in the swarm system. 
	%
	%
\end{itemize}

The traditional approach that relies on separated link for each UAV is not feasible for UAV swarms any more, since simultaneous package transmission will overwhelm the communication system as the number of UAVs increased. Recently, many solutions based on IEEE 802.11 wireless LANs using infrastructure or ad-hoc mode have been proposed \cite{yanmaz2013achieving,chung2016live}. However, most of them are used for short distance scenarios (e.g. in the order of 300m \cite{yanmaz2013achieving}) and even indoor conditions. This restricts the working radius of swarm systems. 
In comparison, radio systems  provide low latency link and more than 10 KM range, and the mesh network and point to multipoint network are supported by current products of radio systems.  Nevertheless, one of the drawbacks is that the bandwidth is relatively small and stays in the order of 100 kbps.

In order to satisfy with the requirements (mentioned above) of autonomous UAV swarms while leveraging the advantages of different wireless techniques, our communication infrastructure consists of 3 types of channels, as shown in Fig.~\ref{fig:comm}.
Firstly, a latency-sensitive channel based on radio systems that is used for command and coordination.
We also propose a custom protocol for exchanging latency-sensitive messages based on mavlink \cite{meier2013mavlink}. By using this custom protocol, messages including location, altitude, commands are simplified and transmitted over this channel.
Secondly, a latency-insensitive channel is used for high-data-rate transmissions, and images and videos are transmitted over this channel.
%
%
Thirdly, a manual control channel based on a radio system that is used for remote control. Note that we rely sole on the first two links for autonomous control of UAV swarms. For the purpose of protecting UAVs when accident happens, a remote control link to each individual UAV is needed.

Besides, the communication layer also includes the communication management software module for transferring messages effectively. More specifically, it is in charge of encapsulating and dis-encapsulating the messages, since the protocols used intra- and inter-aircraft are different. Besides, with the increase in the scale of swarms, optimizing the swarm communication  is also becoming a desire. Application-level optimization approaches such as priority-based traffic scheduling \cite{huang2013priority} and congestion control\cite{rajesh2016congestion} can be included here.

\section{The Human Interaction Layer}\label{sec:gcs}

The Human Interaction layer is in charge of providing the interaction interfaces for the operator to control the swarm system. There are mainly two kinds of interfaces: monitoring interfaces and commanding interfaces. The monitoring interfaces are used for visualizing the situation of the UAV swarm system including the status of UAVs, the sensed data, the environment, etc. And the commanding interfaces are used for sending commands to the UAV swarm system for accomplishing the required missions. Since the UAV swarms may include a large number of UAVs, traditional ground control station (GCS) such as QGroundControl\cite{zurich2013qgroundcontrol} and Paparazzi System\cite{gati2013open}, which provide supervision and control in the flight control level, is not suitable for UAV swarms. This is because the cognition as well as the operating workload is too heavy and will overwhelm the operators. To this end, based on our previous experience of field experiments, we summarize the attributes that the GCS for UAV swarms should have as follow. 

Monitoring Interfaces:
\begin{itemize}
	\item Information about UAV status and sensed data needs to be analyzed pro-actively before representation and the interested results should be highlighted. 
	\item Displays, panels and information entries can be activated or deactivated selectively.  
\end{itemize}

Commanding Interfaces:
\begin{itemize}
	\item Voice assistant can be used for announcing current situation and sending commands.   
	\item High-level commands (e.g. search targets,  track targets, develop formation) should be arranged on the main panel. However, this relies heavily on the autonomy level that the swarm system can reach. 
	\item A workflow pattern of sending a sequence of mission commands can be adopted. High-level commands can be assembled arbitrarily by operators. GCS sends the commands one by one once the acknowledgment previous command is received, or triggers the failsafe policies once errors are detected.
\end{itemize}

Basically, different forms of GCS can be deployed based on the environmental condition, e.g. tablet, vehicles, ships, etc. Besides, due to the convenience for GCS to obtain resources such as computing capacities, power supply, high bandwidth network comparing to the aircraft, we can leverage the high performance servers or even the Data Center facilities as the back end for GCS to perform complex computation and large scale data storage. In this way, tasks such as panoramic mosaic and three-dimensional modeling, which is time-consuming in traditional view, can be completed in real-time.


\section{Experiments and results}\label{sec:results}


In order to evaluate the performance of the proposed architecture, we have built a prototype swarm system based on the proposed architecture and conducted a set of field experiments with five square kilometers. In the following subsections, the experimental set-up as well as the experimental methodology will be presented.

\subsection{Experimental Set-up}

In this section, the experimental set-up utilized for real flights is described in detail. The experimental set-up mainly consists of aerial platforms, integrated on-board control system and the ground control station. 

\subsubsection{Aerial platform}

We use two types of UAVs in the experiments. The first type is a fixed-wing UAV and has a cruise speed of 19 m/s.  The wing span of the vehicle is 1.80 meters and the body length is 1.22 meters. The total weight of the vehicle is 1.1 kilograms and maximum take-off weight is 4.6 kilograms.  The second type is a tilt-rotor UAV and has a cruise speed of 23 m/s.  The wing span of the vehicle is 1.80 meters and the body length is 1.600 meters. The total weight of the vehicle is 1.2 kilograms and maximum take-off weight is 6.89 kilograms. Both of these two UAVs use lithium battery as power supply and have a nominal endurance of 1 hour. 



\begin{figure}
	\centering
	\includegraphics[width=0.25\textwidth]{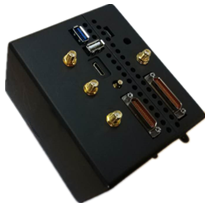}
	\caption{Integrated on-board control system box.}
	\label{fig:hardwarebox}       
\end{figure}

\subsubsection{Integrated on-board control system}

For the purpose of designing a lightweight and miniaturized system, we propose an integrated  on-board control system box that integrates the processing boards, perceptional devices, communication payloads, circuitry and cooling devices into a compact container. This box has a weight of 450g and a size of 108mm(length)$\times$108mm(width)$\times$110mm(height), as shown in Fig.~\ref{fig:hardwarebox}. More specifically, there are three layers in the hardware architecture. The first layer is mainly composed of a custom printed circuit board which is in charge of the power distribution and providing wiring interfaces. The second layer is mainly composed of an autopilot system which is connected to the propulsion system, servos, various sensors, etc. The third layer is mainly composed of a high performance processing board which is connected to telemetry devices and communication payloads. 

By leveraging the computation power of  the processing board, we deploy Linux system and utilize the Robot Operating System (ROS) to implement the on-board software. The functional modules that described in Section \ref{sec:highlevel} are implemented as ROS Nodes, which use the publish-subscribe and client-server mechanisms for inter-nodes communication. 

\begin{figure*}
	\centering
	\subfigure[Vehicle-mounted GCS]{
		\label{fig:gcs:sub1} 
		\includegraphics[width=0.29\textwidth]{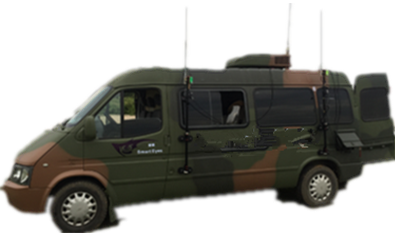}}
	\hspace{0.1in}
	\subfigure[Multi-screen GCS]{
		\label{fig:gcs:sub2} 
		\includegraphics[width=0.23\textwidth]{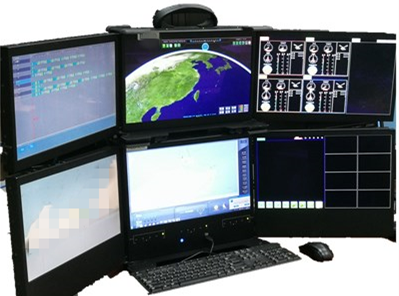}}
	\hspace{0.1in}
	\subfigure[Tablet GCS]{
		\label{fig:gcs:sub3} 
		\includegraphics[width=0.23\textwidth]{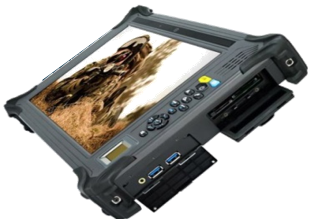}}
	\caption{Various forms of ground control stations.}
	\label{fig:gcs} 
\end{figure*}

\subsubsection{Ground control station}

Fig.~\ref{fig:gcs:sub1} shows the proposed vehicle-mounted ground control station. This ground control station mainly consists of interaction interfaces, communication devices, distributed computing servers and UPS power supply. More specifically, the interaction interfaces can visualize the UAV status, sensed data, the geographical environment as well as diagnosis information (e.g. communication status, wind status). Besides, the operator can command the swarm system through the interaction interfaces. Moreover, in order to satisfy with different requirements of mobility and portability, we also propose other forms of ground control stations as shown in Fig.~\ref{fig:gcs:sub2} and Fig.~\ref{fig:gcs:sub3}. The tablet GCS is used for UAV pilots or high authorities that want to monitor the status of the swarm system. The multi-screen portable GCS is used in the environment where car is not convenient to drive (e.g.  mountainous and lake area). 

Based on the experimental set-up described above, we have built a prototype swarm system and conducted a set of field experiments. In the following subsections, we will evaluate the performance of the proposed architecture.

\subsection{Versatility of the Proposed Architecture}

In this section, we evaluate the versatility of the proposed architecture. To this end, we conduct a set of field experiments in which the prototype UAV swarm system operates multiple missions based on the proposed architecture. Here, we choose the missions of formation flight and target recognition and  tracking as examples. 

\subsubsection{Formation flight}

\begin{figure*}
	\centering
	\includegraphics[width=0.98\textwidth]{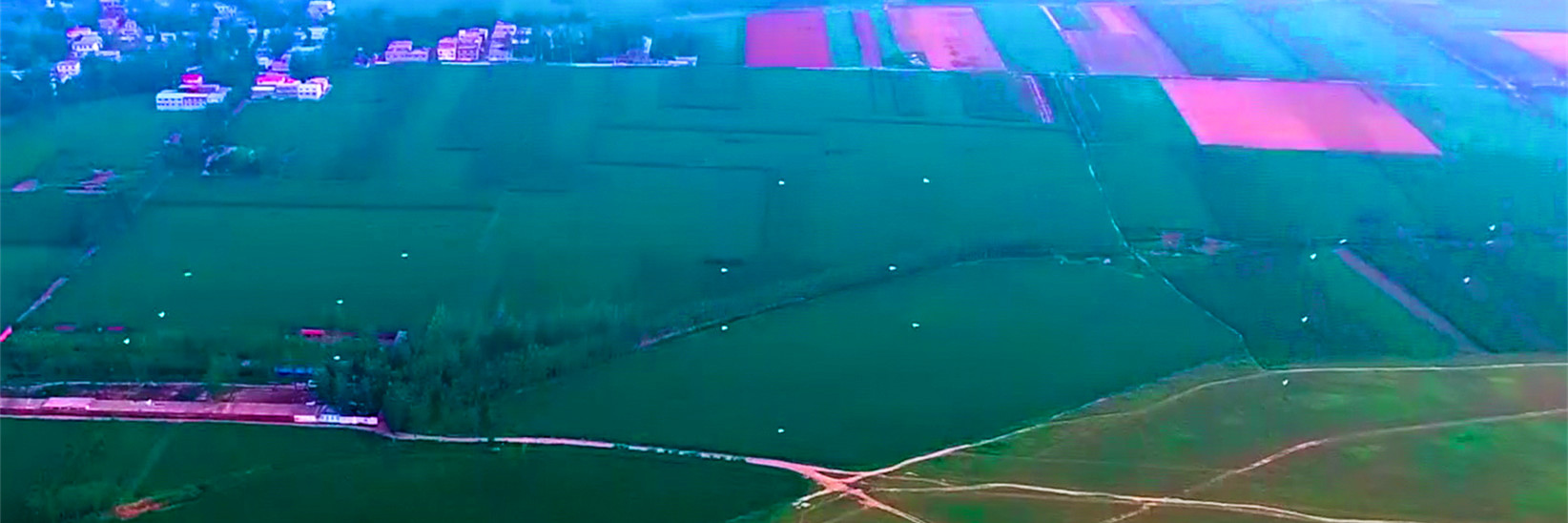}
	\caption{A snapshot of the 21-UAV formation flight experiment. There are three groups. The seven UAVs in each group forms two vertical lines.}
	\label{fig:21}       
\end{figure*}

\begin{figure*}
	\centering
	\subfigure[Triangle pattern]{
		\label{fig:reconfig:sub1} 
		\includegraphics[width=0.32\textwidth]{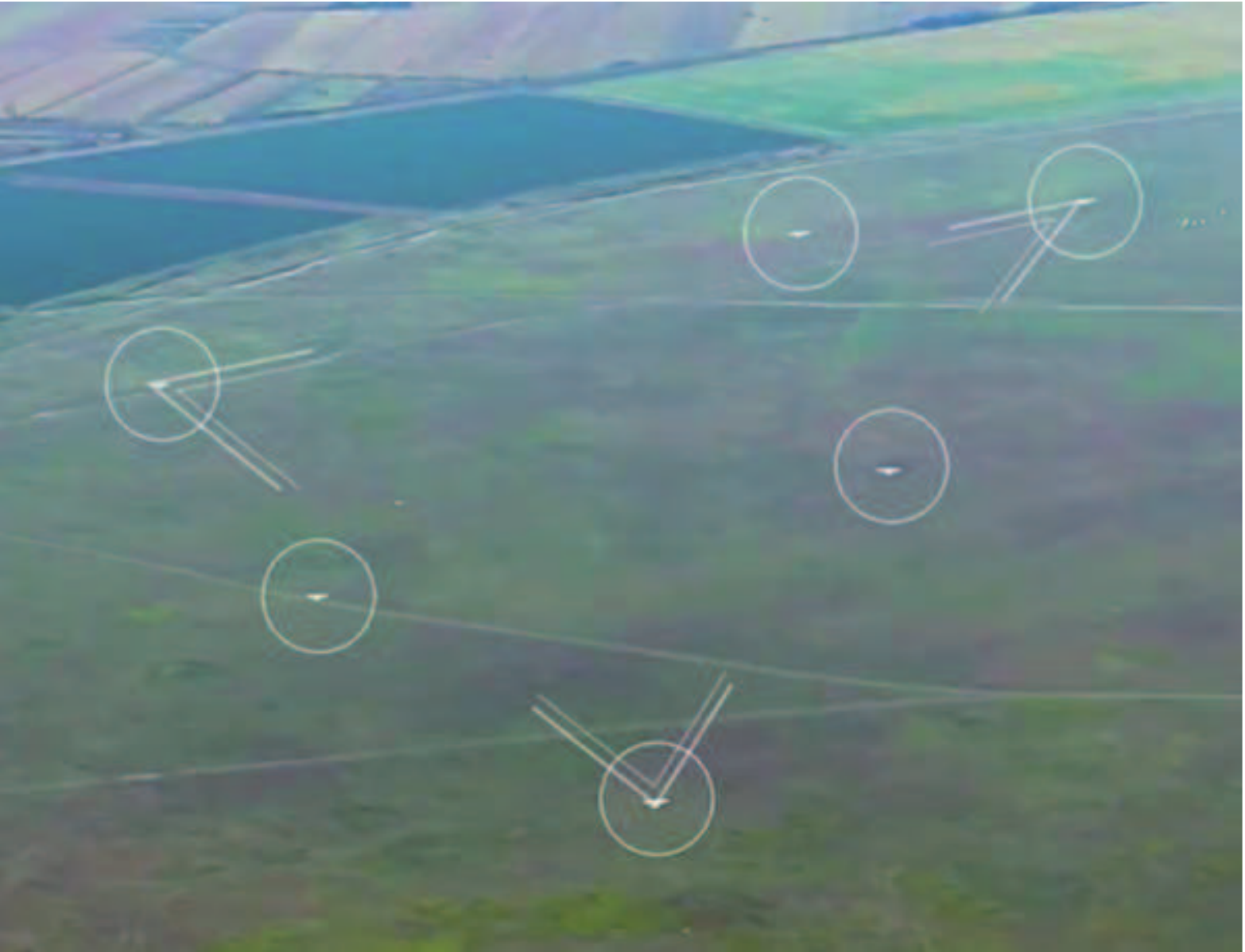}}
	\subfigure[V pattern]{
		\label{fig:reconfig:sub2} 
		\includegraphics[width=0.32\textwidth]{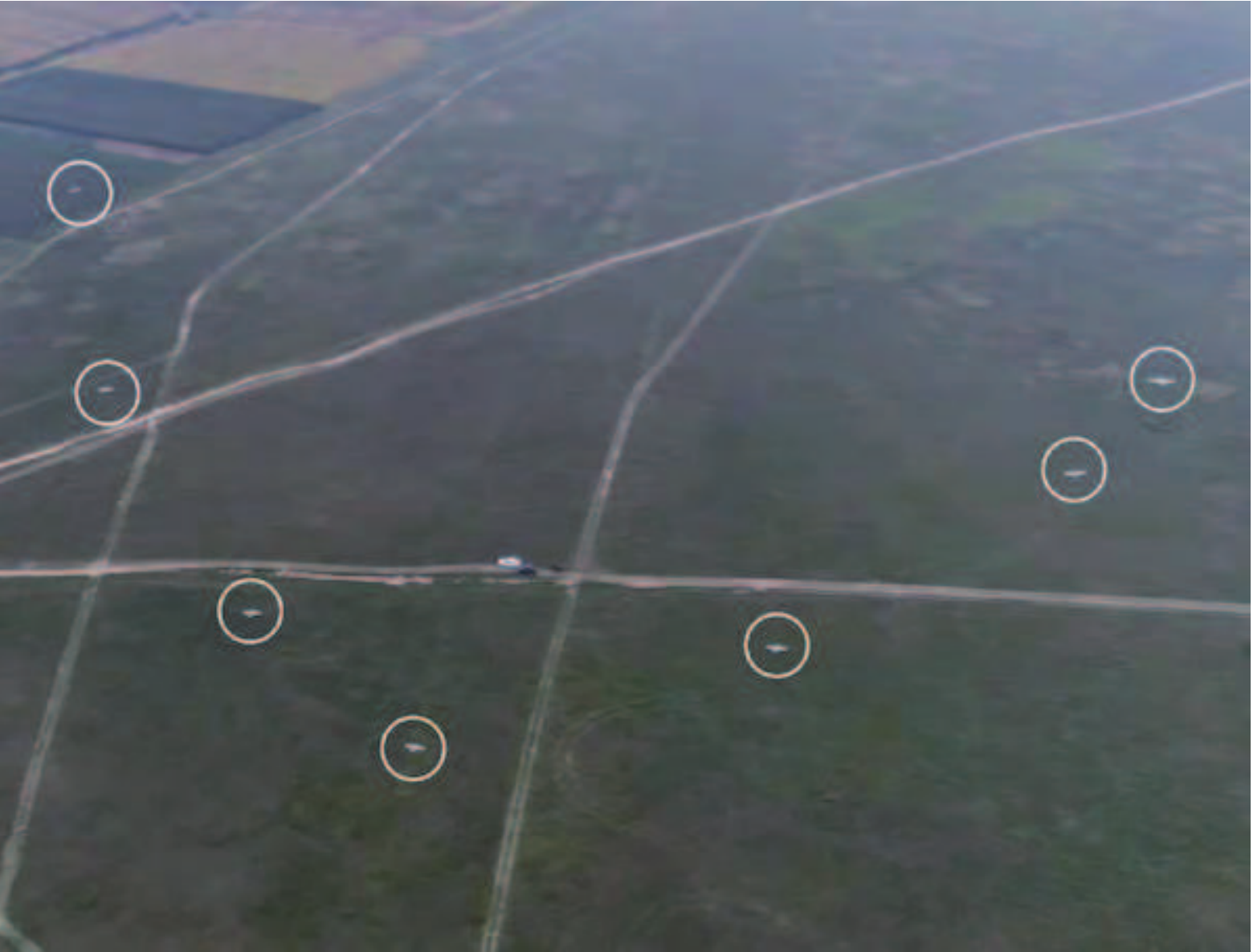}}
	\subfigure[Line pattern]{
		\label{fig:reconfig:sub3} 
		\includegraphics[width=0.32\textwidth]{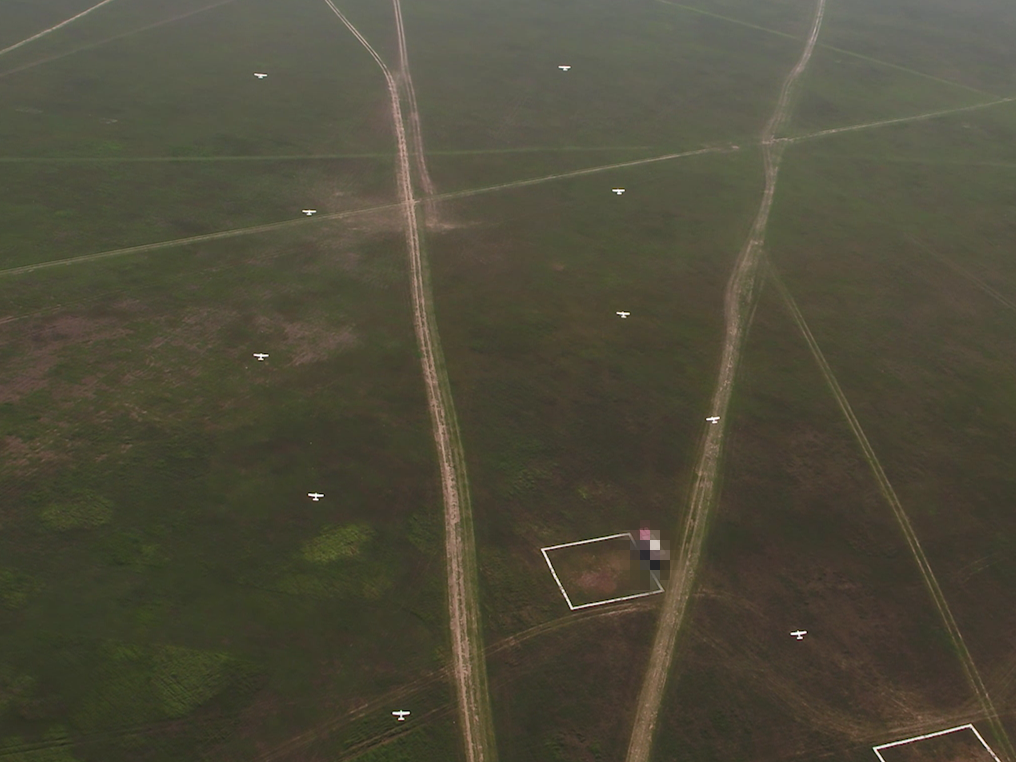}}
	\caption{Formation pattern reconfiguration.}
	\label{fig:reconfig} 
\end{figure*}

Fig.~\ref{fig:21} shows a snapshot of the 21-UAV formation flight experiment. In this experiment, the UAVs are divided into three groups. In each group, there are two leader UAVs and five follower UAVs. 
The leader UAVs execute the coordinated path following guidance control law proposed in Section \ref{sec:guidance}, while the follower UAVs execute the leader-follower coordination guidance control law. As shown in the figure, the seven UAVs in each group forms two vertical lines. 

We have also performed formation pattern maintaining and reconfiguration during the experiment. Fig.~\ref{fig:reconfig} shows examples of different formation patterns that have performed in the experiment. Note that we adopt different strategies of formation pattern reconfiguration for leader and follower UAVs, respectively. For the leaders, the reconfiguration is achieved through coordinated path planning and following. With respect to the followers, we change their relative positions corresponding to their leaders while switching the formation pattern. 
\subsubsection{Target recognition and tracking}



\begin{figure*}
	\centering
	\subfigure[Cooperative Tracking with three UAVs]{
		\label{fig:track:sub1} 
		\includegraphics[width=0.32\textwidth]{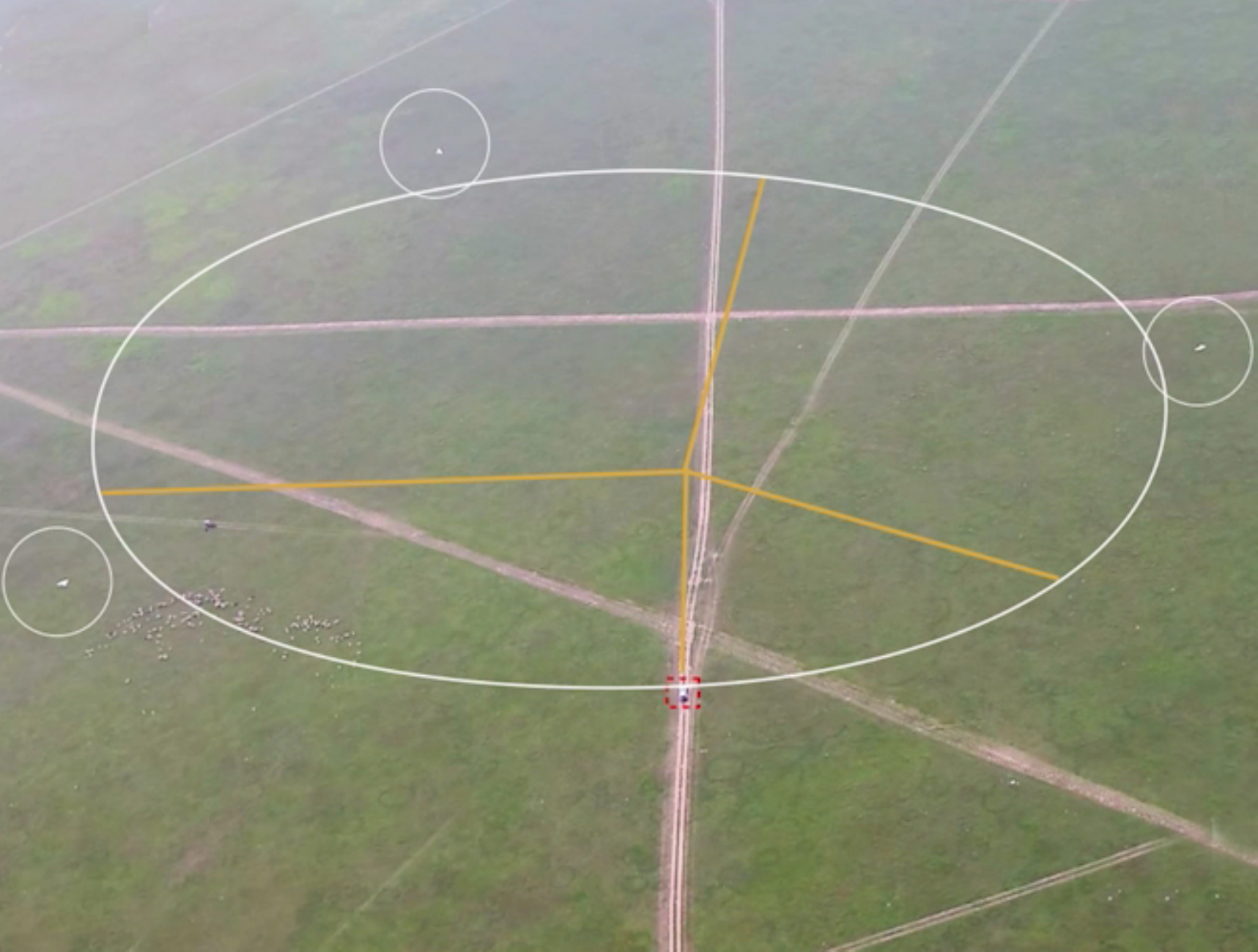}}
	\subfigure[Vision from an UAV]{
		\label{fig:track:sub2} 
		\includegraphics[width=0.32\textwidth]{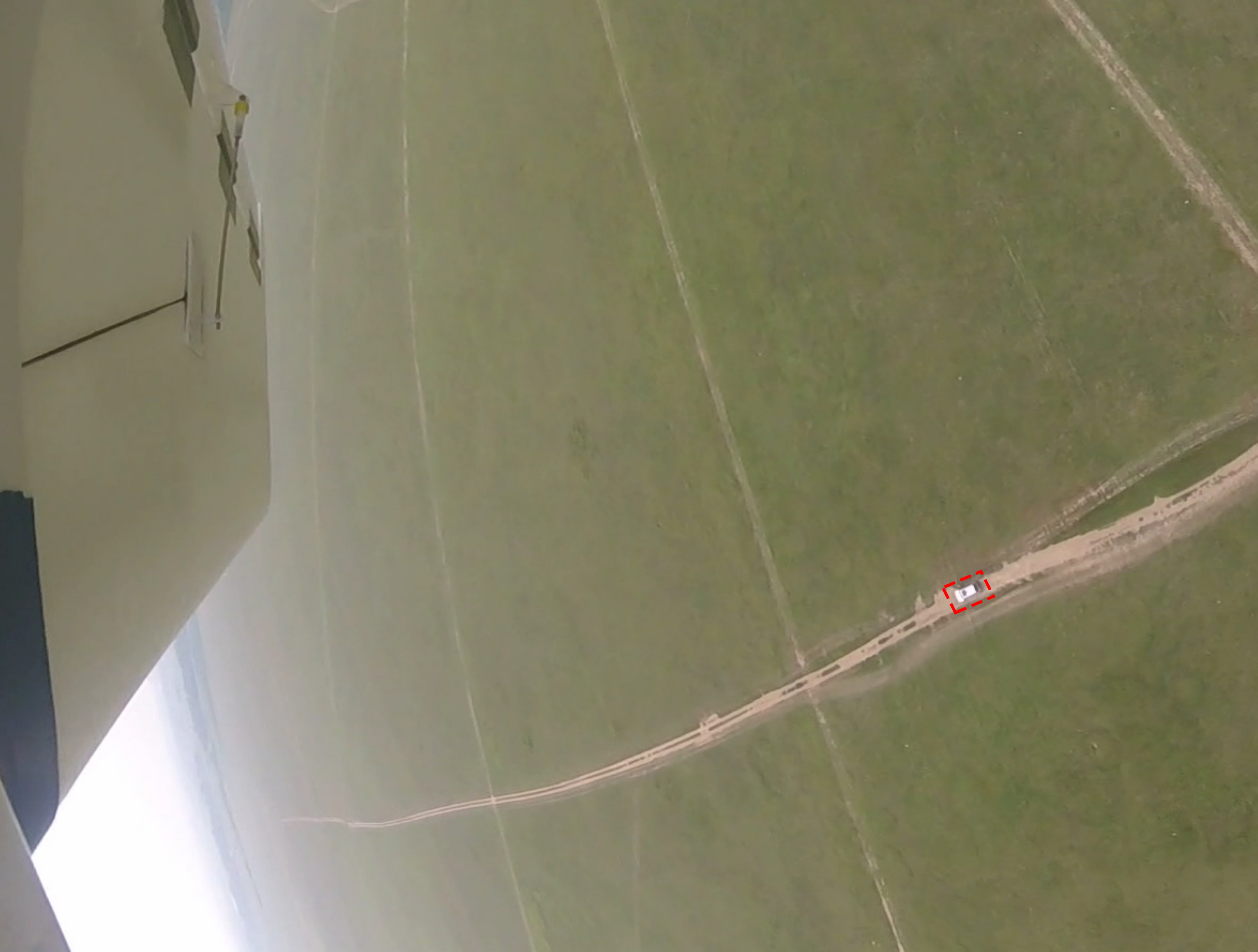}}
	\subfigure[Tracking under the case of shelter]{
		\label{fig:track:sub3} 
		\includegraphics[width=0.32\textwidth]{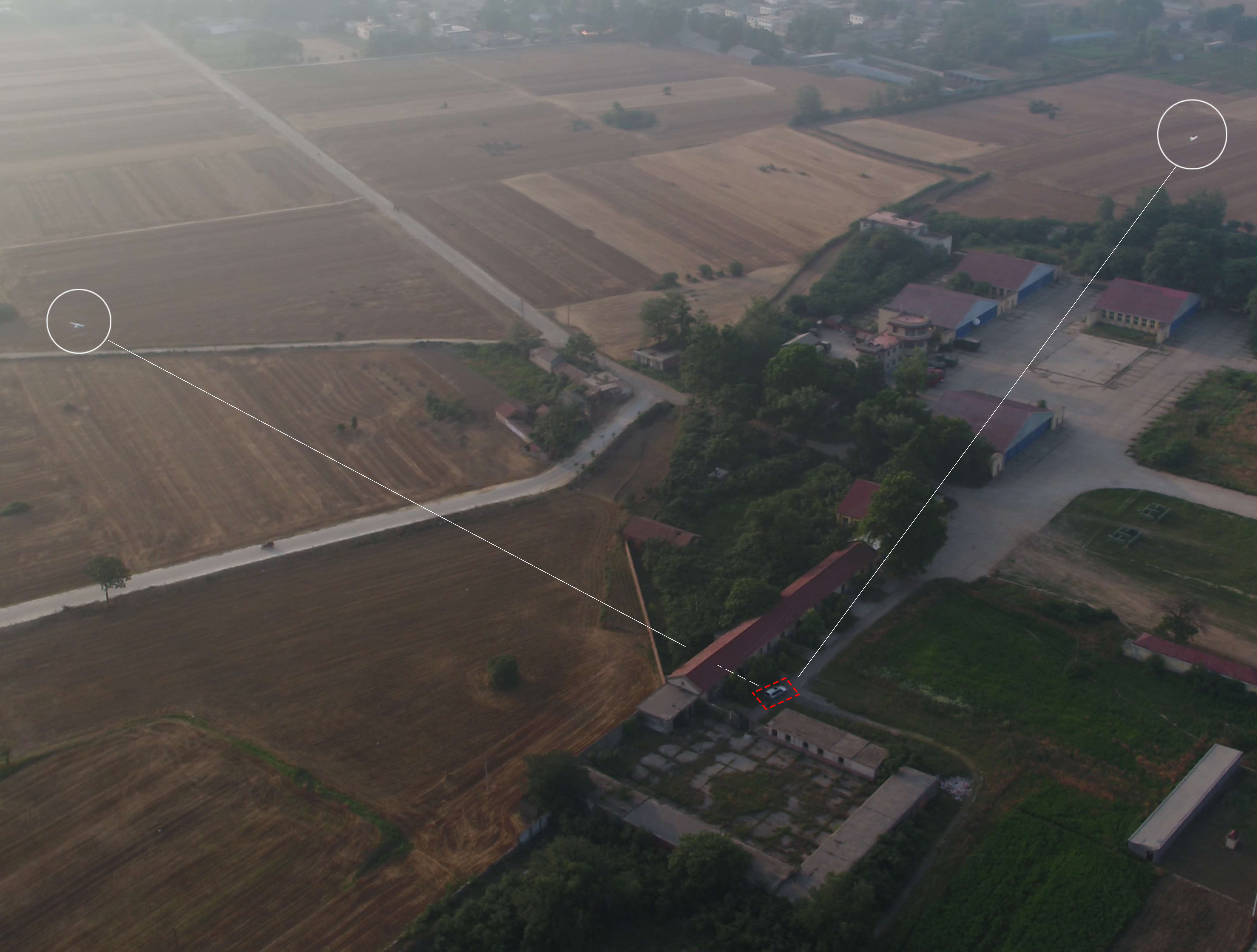}}
	\caption{Target recognition and tracking.}
	\label{fig:track} 
\end{figure*}

\begin{figure*}[!t]
	\centering
	\begin{minipage}[c]{0.48\textwidth}
		\centering
		\includegraphics[height=2.0in]{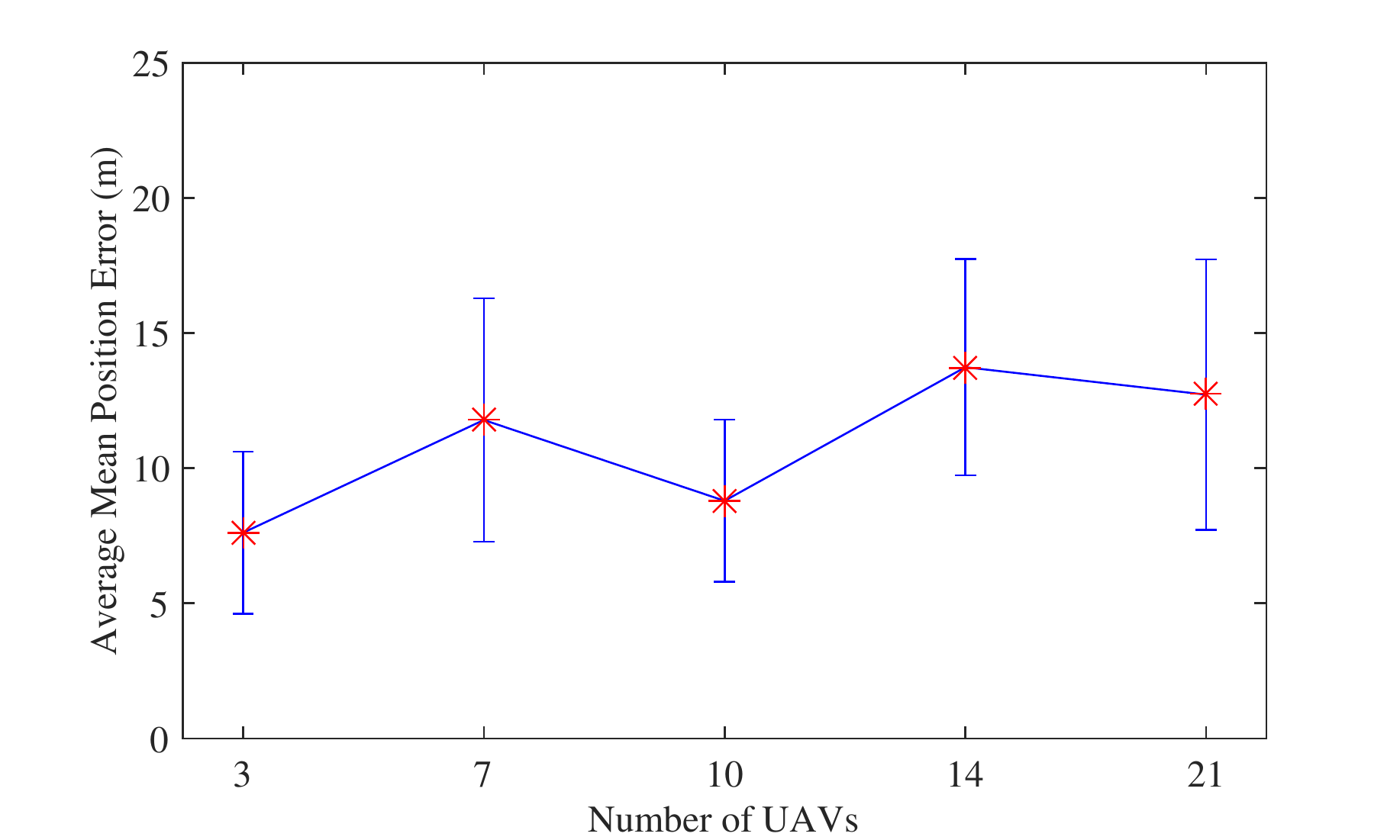}
	\end{minipage}
	\hspace{0.02\textwidth}
	\begin{minipage}[c]{0.48\textwidth}
		\centering
		\includegraphics[height=2.0in]{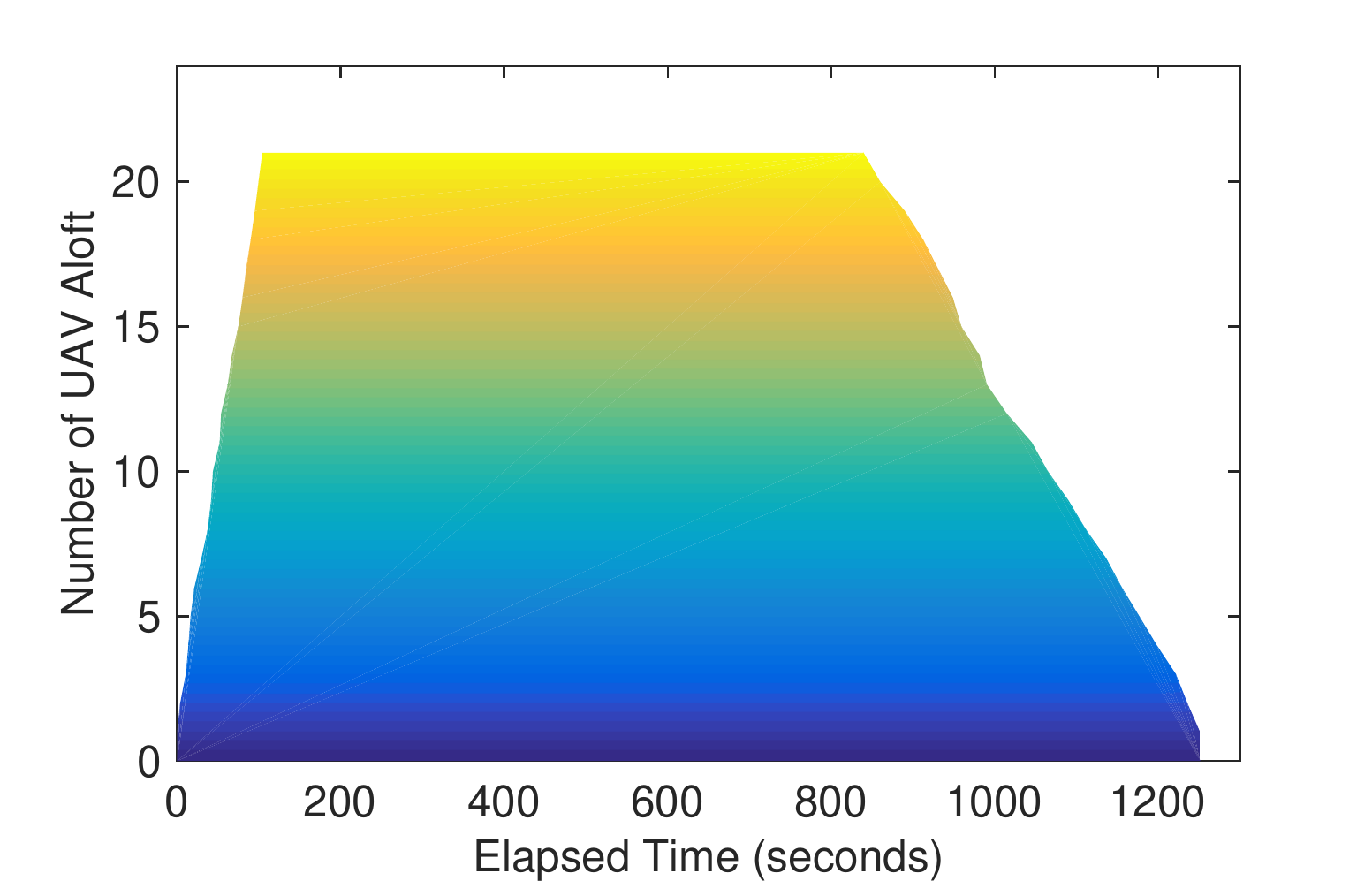}
	\end{minipage}\\[3mm]
	\begin{minipage}[t]{0.48\textwidth}
		\centering
		\caption{The average mean position error for formation flights with increasing number of UAVs using the proposed architecture.}
		\label{fig:ampe}
	\end{minipage}
	\hspace{0.02\textwidth}
	\begin{minipage}[t]{0.48\textwidth}
		\centering
		\caption{Illustration of the evolution of the 21-UAV flight experiment, showing the number of UAV aloft with the time elapsed since the launching command is sent and depicting the sequential launch and landing of all UAVs.}
		\label{fig:aloft}
	\end{minipage}
\end{figure*}

Fig.~\ref{fig:track:sub1} shows a snapshot of the target recognition and tracking experiment with three UAVs. More specifically, each UAV captures videos through its electro-optical pod at runtime. UAVs detect the target through the target detection and recognition algorithm. Once a target is detected, by applying target localization and multi-UAV data fusion algorithm, UAVs can obtain the target information and update the trajectory estimation of the target according to the runtime state. Fig.~\ref{fig:track:sub2} shows a snapshot of the vision from one of the UAVs.  In terms of tracking the target, we adopt cooperative standoff target tracking guidance approach in the experiment. Three UAVs fly a circular orbit around the target with a radius of 100 meter, and each UAV keep an angular spacing of 120 degrees with others. This can achieve the surveillance of  multiple angles, which is useful in the condition that the target is keep out by the shelter such as buildings and trees. As shown in Fig.~\ref{fig:track:sub3}, the target is sheltered by the building that one of the UAV cannot get a vision of the target.

Admittedly, due to the time and manpower limitation, we only demonstrate two different missions in the experiments. However, we want to stress that this paper is the first work, to the best of our knowledge,  which demonstrates formation flight and target tracking missions with an integrated architecture for fixed-wing UAV swarms through field experiments. We believe that other mission can also be supported in the proposed architecture through extension.



\subsection{Scalability of the Proposed Architecture}


In order to evaluate the scalability of the proposed architecture, we access the swarm system's behavior while increasing the scale of the swarm system.
We take the formation flight as an example, and evaluate the performance of formation following behavior through a set of field experiments with increasing number of vehicles. And we use AMPE \cite{navarro2009proposal}, the Average Mean Position Error over all swarm members as well as the time average, as criteria:

\begin{eqnarray}
AMPE = \frac{1}{T_k}\sum_{t=1}^{T_k} MPE(t), \\
MPE(t) = \frac{1}{N}\sum_{i=1}^{N}\left\|P_{i}^{D}(t)-P_{i}^{R}(t)\right\|,
\end{eqnarray}
where $N$ is the number of vehicles performing the formation flight. $T_k$ is the number of timestamps during the duration of the formation flight. $P_{i}^{D}(t)$ and $P_{i}^{R}(t)$ denote the desired and real position of UAV $i$ at timestamp $t$, respectively. Fig.~\ref{fig:ampe} shows the average mean position error for formation flights with 3, 7, 10, 14, 21 UAVs using the proposed architecture. It is clear that as the number of the vehicles increases, the average mean position error for formation flights hovers around 10 meters. As far as we know, many existing work are evaluated though field experiments outdoor, however, most of them are small scale (e.g. two to five). Therefore, it can be seen that the proposed architecture shows decent scalability while increasing the scale of the swarm system. Admittedly, the scalability of the architecture relies heavily on the communication infrastructure. And many solutions such as hierarchical design and multi-cast can improve the scalability of the communication. 
However, this is out of the scope of this paper. 
In this paper, we focus on the architecture design in the level of the swarm system.

\subsection{Launch and landing rate}

Generally speaking, the endurance of miniaturized UAV is limited, which directly effects its mission capability.  In order to reduce the time overhead of deploying UAV swarms into the sky, fast launch and recovery of UAV swarms is a basis desire. However, compared with a single UAV, launching and recovering a swarm of UAVs are much more challenging, especially for fixed-wing UAV. Rational planning and collision avoidance need to be considered since many UAVs gather around a limited airspace when launching and recovering. In our proposed swarm system, we launch and recover UAVs one by one through short distance taxing. The average taxing distances for launching and recovering are 20 meters and 50 meters, respectively. Fig.~\ref{fig:aloft} illustrates the evolution of a 21-UAV flight experiment, showing the number of UAV aloft since the launching command is send. It is clear that this flight lasts for 1257s, and at 110.43s all 21 UAV is aloft. Similar to~\cite{chung2016live}, we use the mean time between launches as the criterion to measure the launch rate of UAV swarms, which is,

\begin{equation}\label{eq:launch}
LaunchRate = \frac{1}{N}\left(\tau_{launch}^{1}+\sum_{i=2}^{N}\left(\tau_{launch}^{i} -\tau_{launch}^{i-1}\right)\right),
\end{equation}
where $\tau_{launch}^{i}$ is the time between the $i^{th}$ UAV is aloft and the launching command of the swarm system is sent. Accordingly, we can obtain the $LaunchRate$ of our swarm system is $ 5.25$ seconds. This is significantly faster than the launch rate in~\cite{chung2016live}, which stays at $33.5$ seconds. 

Similarly, we can measure the landing rate by using the criterion as Equation \ref{eq:land},

\begin{equation}\label{eq:land}
LandRate = \frac{1}{N}\left(\tau_{land}^{1}+\sum_{i=2}^{N}\left(\tau_{land}^{i} -\tau_{land}^{i-1}\right)\right),
\end{equation}
where $\tau_{land}^{1}$ is the time between the $i^th$ UAV is landed and the landing command of the swarm system is sent. Accordingly, we can obtain the $LandRate$ of our swarm system is $ 21.71$ seconds. Although Chuang et al. \cite{chung2016live} have not provided the precise land rate for their system,  we can see from Fig.~7 in the paper that it takes more than 25 minutes for landing 50 UAVs. Hence, we take the average of this duration to estimate the Land rate in \cite{chung2016live}, which is 30s. Therefore, the land rate of our swarm system is also faster than the land rate in \cite{chung2016live}.
%


\section{Conclusions}\label{sec:conclusion}
In this paper, we presented a multi-layered and distributed architecture for mission oriented miniature fixed-wing UAV swarms and a holistic view of a swarm system including hardware, software, communication and human interaction is detailed. Compared to exiting solutions, the proposed architecture divides the overall system to five layers: low-level control, high-level control, coordination, communication and human interaction layers, and many modules with specified functionalities. 
As a result, not only the difficulty of developing a large system can be reduced, but also the versatility of supporting diversified missions can be ensured. Besides, the proposed architecture is fully distributed that each UAV performs decision-making autonomously, 
thereby achieving better scalability. Moreover, different kinds of aerial platforms can be feasibly extended by using the control allocation matrix and the integrated hardware box. We conducted field experiments, including autonomous launch, formation flight, target recognition and tracking and landing of 21 fixed wing UAVs, which evaluated the scalability and versatility of the proposed architecture. The experimental results also show that the launch rate of the prototype system based on the proposed architecture outperforms the state-of-the-art work. 

There are some interesting future works in this direction. One possible future work is to extend other collective behaviors and coordination missions, such as search and rescue, forest fire control and cooperative surveillance or mapping, to a larger scale of UAV swarm.
Another future work is to provide an open-source software framework that serves as a testbed for promoting  advances in UAV swarming community. 

\section*{Acknowledgment}

The authors would like to express their deepest gratitude to the SWARM TEAM of the NUDT. Without their hard work, the field experiments cannot be done.

\ifCLASSOPTIONcaptionsoff
  \newpage
\fi




\bibliographystyle{IEEEtran}
\bibliography{uav}
\end{document}